\documentclass[fleqn,10pt]{wlscirep}
\usepackage[utf8]{inputenc}
\usepackage[T1]{fontenc}
\title{Mid-infrared Single-photon Detection Using Superconducting NbTiN  Nanowires with Sub-15 ps Time Resolution in a Gifford-McMahon Cryocooler}

\author[1,2,$\dagger$]{Jin Chang}
\author[2]{Johannes W. N. Los}
\author[2]{Ronan Gourgues}
\author[3]{Stephan Steinhauer}
\author[2]{S. N. Dorenbos}
\author[1]{Silvania F. Pereira}
\author[1]{H. Paul Urbach}
\author[3]{Val Zwiller}
\author[1]{Iman Esmaeil Zadeh}

\affil[1]{Optics Research Group, ImPhys Department, Faculty of Applied Sciences, Delft University of Technology, Delft 2628 CJ, The Netherlands.}
\affil[2]{Single Quantum B.V., Delft 2628 CJ, The Netherlands.}
\affil[3]{KTH Royal Institute of Technology, Department of Applied Physics, Albanova University Centre, Roslagstullsbacken 21, 106 91 Stockholm, Sweden}

\affil[$\dagger$]{\emph{Corresponding author:j.chang-1@tudelft.nl}}

\begin{abstract}
Shortly after their inception \cite{2001}, superconducting nanowire single-photon detectors (SNSPDs) became the leading  quantum light detection technology \cite{phy_and_application}. With the capability of detecting single-photons with near-unity efficiency \cite{98simit,99delft,98nist}, high time resolution \cite{7.7ps,3ps}, low dark count rate \cite{low-dcr}, and fast recovery time \cite{munzberg2018superconducting}, SNSPDs outperform conventional single-photon detection techniques. However, detecting lower energy single-photons (<0.8\,eV) with high efficiency and low timing jitter has remained a challenge. To achieve unity internal efficiency at mid-infrared wavelengths, previous works \cite{Mo80Si20,WSi10um} used amorphous superconducting materials with low energy gaps at the expense of reduced time resolution (close to a nanosecond \cite{1.5-6um}), and by operating them in complex mK dilution refrigerators. In this work, we provide an alternative approach with SNSPDs fabricated from 5-9.5\,nm thick NbTiN superconducting films and devices operated in conventional Gifford-McMahon (GM) cryocoolers. By optimizing the superconducting film deposition process, film thickness and nanowire design, our fiber-coupled devices achieved > 70\% system detection efficiency (SDE) at 2\,\textmu m  and sub-15\,ps timing jitter. Furthermore, detectors from the same batch demonstrated unity internal detection efficiency at 3\,\textmu m and 80\% internal efficiency at 4\,\textmu m, paving the road for an efficient mid-infrared single-photon detection technology with unparalleled time resolution and without mK cooling requirements. We also systematically studied the dark count rates (DCRs) of our detectors coupled to different types of mid-infrared optical fibers and black-body radiation filters. This offers insight into the trade-off between bandwidth and dark count rates for mid-infrared SNSPDs. To conclude, this paper significantly extends the working wavelength range for SNSPDs made from polycrystalline NbTiN  to 1.5-4\,\textmu m, and we expect quantum optics experiments and applications in the mid-infrared range to benefit from this far-reaching technology. 
\end{abstract}
\begin{document}

\flushbottom
\maketitle

\section*{Introduction}
Detecting light at the single photon level has enabled novel scientific and industrial applications in recent decades \cite{phy_and_application}. Specifically, near- and mid-infrared detection are crucial for areas such as infrared fluorescence and spectroscopy \cite{mir-spec,mir-fluorescence,qd-snspd}, semiconductor and industrial production monitoring \cite{cmos-test,industry-monitor},  planetary soil studies \cite{mercury}, remote light detection and ranging \cite{2.3lidar} as well as two-photon entanglement and interference \cite{2-photon} experiments. However, since photon energy is inversely proportional to wavelength, detecting long wavelength photons is intrinsically more challenging than detecting shorter wavelength photons. Generally, Si-based detectors can be used for infrared detection but suffer from a low cut-off wavelength, typically around 1.1\,\textmu m \cite{infrared-review1}, making them inefficient for long wavelength photon detection. Si:Sb based impurity band conduction detectors show mid-infrared light detection capability but not at the single-photon level \cite{huffman1992si}. Similarly, narrow-bandgap photoconductive semiconductors, like HgCdTe, InAs and InGaAs detectors suffer from low efficiency, large dark counts and poor time resolution \cite{infrared-review2}. In contrast, SNSPDs have high detection efficiency \cite{98simit,99delft,98nist}, high detection rates \cite{1.5GHz}, low dark count rates (DCR) \cite{low-dcr}, unprecedented temporal resolution \cite{3ps,7.7ps} and thus outperform traditional infrared single-photon detectors.
\par            
In 2001, NbN-based SNSPDs were first demonstrated by detecting 810\,nm single-photons \cite{2001}. Subsequently, SNSPDs fabricated on different platforms were explored and developed \cite{phy_and_application}. Although high system detection efficiencies have been realized and reported for the UV \cite{uv}, visible \cite{mm-jin} and near-infrared/telecom \cite{le2016high,98simit,99delft,98nist}, detecting single-photons beyond 1550\,nm with high efficiency and time resolution has remained a challenge \cite{taylor2021mid}. Early works showed amorphous WSi based SNSPDs could be used for mid-infrared detection. However, these studies employed 4-6\,nm thin superconducting films, resulting in low critical currents which is detrimental to the timing jitter. The reported temporal resolution was close to the nanosecond scale \cite{1.5-6um}. Also, these devices must be operated at sub-Kelvin temperatures, requiring complex dilution refrigerators. NbN-based SNSPDs with ultra-narrow line widths showed sensitivity up to 5\,\textmu m (saturated internal efficiency until 2.7\,\textmu m) \cite{1-5um}. A consequence of squeezing the nanowire width to around 30\,nm makes fabrication challenging and degrades the detectors' time resolution with the reduced critical current. \par 
Alternatively,  our previous work \cite{Ti-sweep} showed that by optimizing the stoichiometry of polycrystalline NbTiN film during reactive magnetron co-sputtering deposition, it is possible to make SNSPDs with strongly saturated efficiency plateaus in the near-infrared region at 2.8\,K operating temperature, and also high performance at visible wavelengths up to 7\,K \cite{4-7}. Also, relatively thick NbTiN superconducting films were used \cite{7.7ps,cleo} to improve our detectors' optical absorption and critical current, therefore enhancing efficiency and time resolution. Building on our previous results, in this work we made SNSPDs from 5, 6.5, 7.5 and 9.5\,nm thick NbTiN films with different nanowire designs. First, by characterizing our SNSPDs using flood illumination, we optimized the meander design in terms of internal detection efficiency. Encouraged by our initial characterization results, we fabricated fiber-coupled SNSPDs and achieved a system detection efficiency of >70\% at 2\,\textmu m in Gifford-McMahon (GM) cryo-coolers (2.4-2.8\,K). Broadband detectors were also demonstrated with >50\% SDE over the entire 1550-2000\,nm range with sub-15\,ps timing jitter. Furthermore, devices made from 7.5 and 6.5\,nm films showed unity internal detection efficiency at 3\,\textmu m and 80\% internal efficiency at 4\,\textmu m. We also systematically studied the DCRs from the detector itself (intrinsic DCRs) and from black-body radiation delivered by different types of fibers as well as coated fibers as a technique to reduce the DCR. These results offer a comprehensive understanding of the origin of dark counts in mid-infrared SNSPD systems. 

\section{SNSPD Fabrication and Measurement Setup}
\begin{figure}[ht]
    \centering
    \includegraphics[width=15cm]{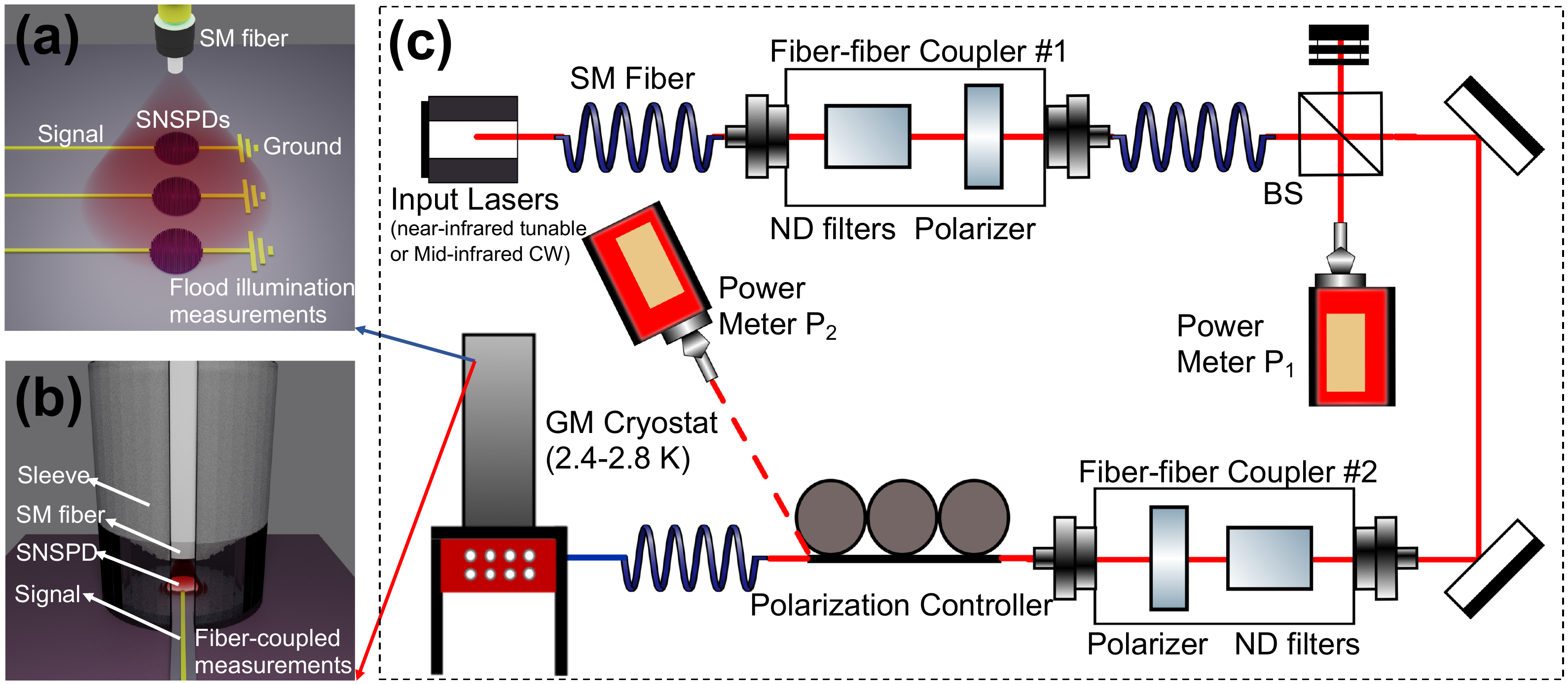}
    \caption{Illustration of \textbf{(a)} detector flood illumination, \textbf{(b)} a fiber-coupled detector, and \textbf{(c)} schematic of the efficiency measurement set-up.  }
    \label{fig:set-up}
\end{figure}

Similar to \cite{Ti-sweep}, we deposited superconducting NbTiN films by a reactive magnetron co-sputtering deposition process. The stoichiometry of the films was controlled by adjusting the sputtering powers on the Ti and Nb targets. Film thickness was determined by a calibrated crystal microbalance and SNSPDs were fabricated as described in \cite{7.7ps}. Our fabricated detectors were either tested under flood illumination (figure \ref{fig:set-up} (a)), or etched into a key-hole die shape and packaged using a standard ferrule and mating sleeves approach \cite{miller2011compact} (figure \ref{fig:set-up} (b)). This coupling method guarantees automatic alignment between detector and optical fibers for accurate system efficiency measurements. Both the flood illumination and fiber-coupled setup are shown in figure \ref{fig:set-up} (c). \par
As shown in figure \ref{fig:set-up} (c), we employ a near-infrared tunable laser (JGR-TLS5, 1260-1650\,nm) and mid-infrared CW lasers with different wavelengths (2000 and 2700\,nm laser from Thorlabs, 3001 and 4013\,nm laser from Nanoplus) as input photon sources. Single mode optical fibers are used to couple the light to the first fiber-to-fiber coupler (containing neutral density filters and a polarizer). A beam splitter is then used to create a reference arm with the majority of the power coupled to a calibrated power meter $P_1$. The signal beam with the lowest power is sent to the second fiber-to-fiber coupler, also containing a polarizer and neutral density filters. A polarization controller is used to tune the polarization state of the light after the second fiber-to-fiber coupler. After recording the light power intensity emerging from the polarization controller with power meter $P_2$, the light is guided to the system to carry out either flood illumination or fiber-coupled measurements. For flood illumination measurements, the input light is heavily attenuated to the single-photon regime. For fiber-coupled device measurements we use the following procedure: we first set the ratio $P_1/P_2$ to 50\,dB by placing ND (neutral density) filters in both fiber-to-fiber couplers, and then add additional ND filters to the first coupler to reach $P_1$ = 10\,nW. In this way, the input photon flux can be back calculated. For example, 10\,nW with 50\,dB attenuation at 2000\,nm corresponds to an input photon flux of $1.006\times10^6$ photons per second. More measurement details can be found in our previous work \cite{99delft}.

\section{Characterization of SNSPDs with Flood Illumination }
In this work, SNSPDs with different nanowire widths (40/60/80\,nm) and diameters (8/9/10\,\textmu m) were fabricated from 5-9.5\,nm thick NbTiN films. For example, 60-120-r4 refers to a meandering nanowire design with 60\,nm wide lines, a pitch of 120\,nm, and 4\,\textmu m radius (see insert in figure \ref{fig3} (b)). 

\begin{figure}[hthp]
    \centering
    \includegraphics[width=17cm]{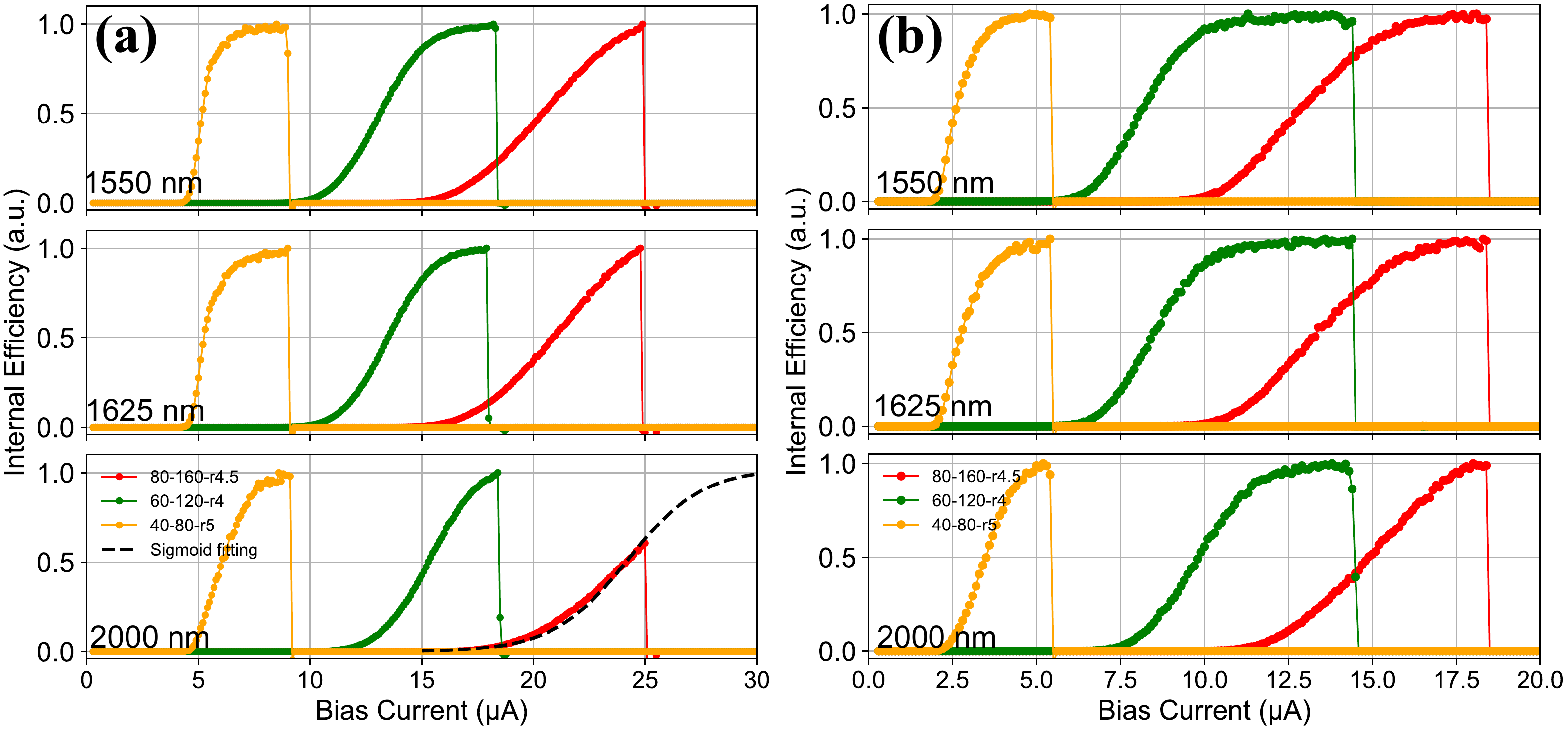}
    \caption{Internal efficiency measurements of SNSPDs fabricated from \textbf{(a)} 9.5\,nm and \textbf{(b)} 7.5\,nm thick NbTiN films.}
    \label{freespace-eff}
\end{figure}
As shown in figure \ref{freespace-eff} (a), at 1550 and 1625\,nm, the 40\,nm width device (yellow), 60\,nm width device (green ) and 80\,nm width device ( red ) all showed saturated internal efficiencies. When the laser wavelength was increased to 2000\,nm, the 40 and 60\,nm wide nanowires (yellow and green) devices maintained saturated internal efficiencies, while the 80\,nm wide nanowire device (red curve) does not reach unity internal efficiency. To obtain greater saturated internal efficiency, one possible solution is to make narrower lines. However, with narrower line width (<40\,nm) the nanofabrication patterning, development, and etching, become more critical. This will, in general, affect the fabrication yield. Alternatively, we sputtered 7.5\,nm thick NbTiN films and made detectors with the same designs and nanofabrication process. As shown in figure \ref{freespace-eff}(b), by using 7.5\,nm thick NbTiN film, all devices with line widths ranging from 40-80\,nm showed unity internal efficiency at 2000\,nm. Detailed performance of devices based on 9.5 and 7.5\,nm films is summarized in table \ref{table:table}. A thinner film leads to lower critical currents (for the same meander design), timing jitter is thus higher because the output pulse has a lower signal to noise ratio \cite{SNR}. The rise time (time interval for signal to go from 20\%-80\% of the pulse amplitude) of the devices on 7.5\,nm NbTiN was longer than for the 9.5\,nm devices and dead-time (width of pulse at level of 1/e of the amplitude) was also slightly longer, this can be explained by the fact that devices made with thinner films have higher kinetic inductance. \par
\makeatletter
\def\thickhline{%
  \noalign{\ifnum0=`}\fi\hrule \@height \thickarrayrulewidth \futurelet
   \reserved@a\@xthickhline}
\def\@xthickhline{\ifx\reserved@a\thickhline
               \vskip\doublerulesep
               \vskip-\thickarrayrulewidth
             \fi
      \ifnum0=`{\fi}}
\makeatother
\newlength{\thickarrayrulewidth}
\setlength{\thickarrayrulewidth}{3\arrayrulewidth}

\begin{table}[ht]
\centering
\begin{tabular}{ p{1.7cm}p{1.7cm}p{2.2cm}p{2.3cm}p{1.7cm} }

\thickhline
 Meander \par Structure&
 $I_c$ (\textmu A)\par 9.5/7.5\,nm &
 Rise-time (ps)\par 9.5/7.5\,nm&
 Dead-time (ns)\par 9.5/7.5\,nm&
 Jitter (ps)\par 9.5/7.5\,nm\\
 \hline
 80-160-r4.5 & 25.0/18.4    &350/375&   9.3/10.6&30/44\\
 60-120-r4   &18.2/14 & 325/335&  11.6/12.6&40/45\\
 40-80-r5    &8.40/6 & 400/425&  38.4/49.2&93/97\\
 
\thickhline
\end{tabular}
\caption{Flood illumination measurement results of 9.5 and 7.5\,nm NbTiN based SNSPDs.}
\label{table:table}
\end{table}

\begin{figure}[!ht] 
    \centering
    \includegraphics[width=17cm]{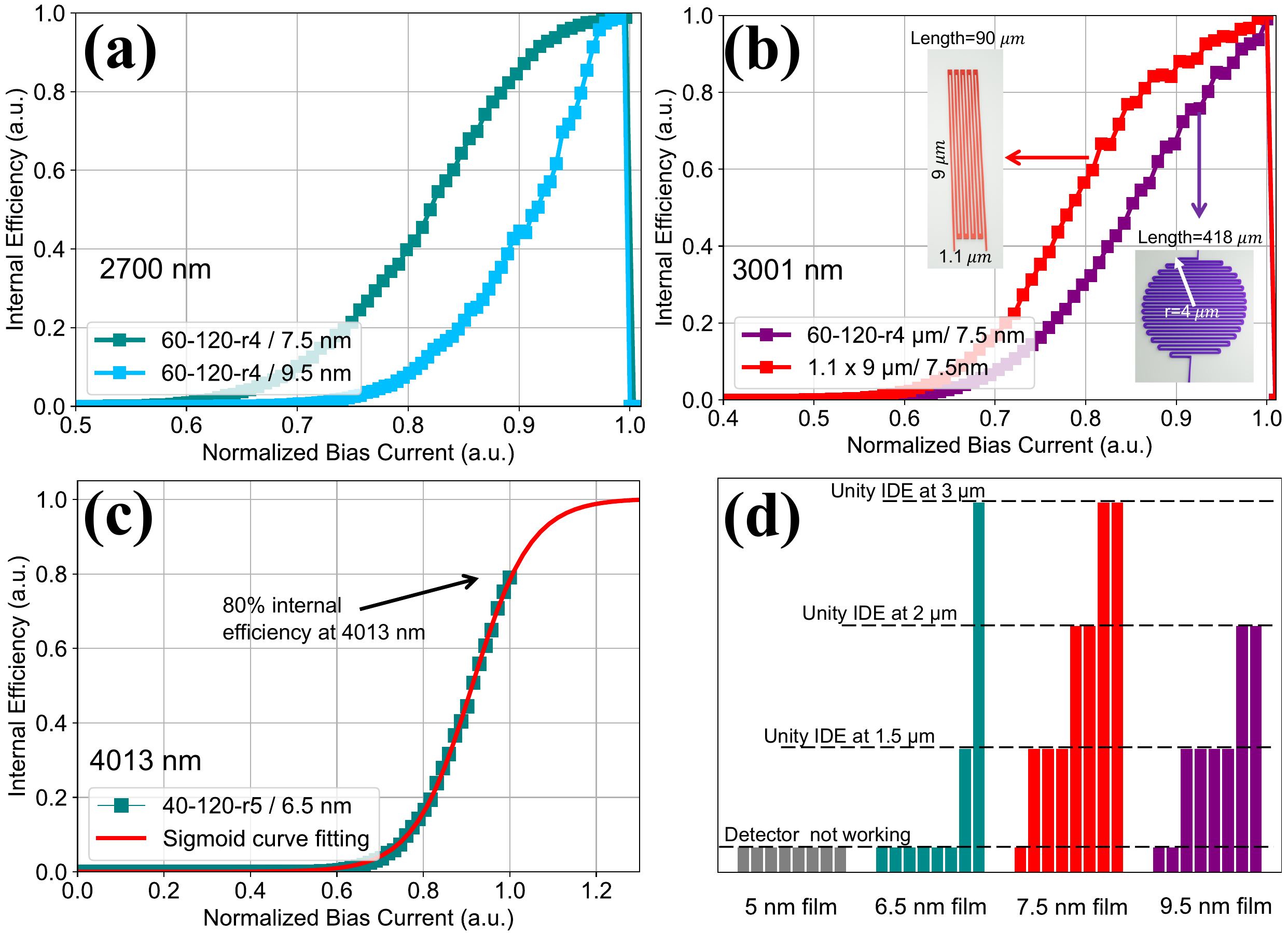}
    \caption{\textbf{(a)} Photon counting rate (PCR) curves at 2700\,nm of 60-120-r4 detectors from 7.5 nm (green) and 9.5\,nm (blue) film, \textbf{(b)} PCR curves at 3001\,nm of 60-120-r4 detector/purple and 1.1$\times$9 \textmu m detector/red from 7.5\,nm film, \textbf{(c)} PCR curves at 4013\,nm of a 40-120-r5 detector from 6.5 nm film, and \textbf{(d}) statistics of device yield made from films with different thicknesses.}
    \label{fig3}
\end{figure}

The above results show that saturated internal efficiency until 2000\,nm can be obtained with detectors made from 9.5/7.5\,nm thick films. In order to explore the internal saturation limit, we carried out longer wavelength flood illumination measurements at 2700 and 3001\,nm for a number of selected detectors. Figure \ref{fig3} (a) shows detectors with 60-120-r4 meander design from both 9.5/7.5\,nm films at 2700\,nm. Both detectors reach unity internal efficiency and detectors from 7.5\,nm film (dark green curve) show stronger saturated internal efficiency than detectors from 9.5\,nm film (light blue curve). This is because by reducing the thickness of the superconducting film, the superconducting energy gap is reduced with the same input photon power, it is easier to break the superconducting state and form a resistive region \cite{thickness}. \par 
Previous measurements at 2700\,nm indicate that detectors made from 7.5\,nm films are still promising for detecting single photons beyond 2700\,nm, we fabricated two types of SNSPDs from a 7.5\,nm NbTiN film and measured their detection performances at 3001\,nm. As shown in figure \ref{fig3} (b), a 'large' meandering nanowire detector design of 60-120-r4 (purple), and a 'small' detector design with line-width 60\,nm, filling factor 50\%, $1.1\times9$ \textmu m (red) were employed. Both detectors show saturated internal efficiency at 3001\,nm and the smaller detector shows superior saturated internal efficiency over the larger one. According to a previous study \cite{inhomo}, the performance of SNSPDs is influenced by the inhomogeneity of the superconducting film. Since the total length of the small detector ($\sim90$ \textmu m) is more than 3 times shorter than the large ($\sim418$ \textmu m), less inhomogeneity can be expected and better detection performance is observed. This shows that by reducing SNSPD's total length, better detection performance can be potentially achieved. In previous works \cite{WSi10um}, the best performing device was a single 10-\textmu m-long line. As a consequence the active area is smaller, which can be increased by using different detector architectures, for example multi-pixel \cite{kilo-pixel} or interleaved nanowire designs \cite{interleaf}. \par

Finally, we evaluated detectors made from even thinner films (5 and 6.5\,nm). In figure \ref{fig3} (c), we demonstrate that a detector (40-120-r5) from 6.5\,nm film achieves 80\% internal efficiency at 4013\,nm (determined using a sigmoid curve fitting). This represents the state-of-the-art mid-infrared  polycrystalline material based SNSPDs. To get a better understanding of the film thickness on the detector performance, we created an overview in figure \ref{fig3} (d). We present the statistics of 32 fabricated SNSPDs from 4 different films. It is clear that 5 and 6.5\,nm films suffer from low yield. 
The detectors made from the 5\,nm films do not work well, because of their low critical current (1-2  \textmu A). The non-working detectors from the 6.5\,nm film did not show unity internal efficiency at 1550\,nm possibly caused by lower film homogeneity of the thin film \cite{7.7ps}. In contrast, 7.5 and 9.5\,nm films show higher yield but detectors from the 9.5\,nm film start to show decreased internal efficiency in the mid-infrared compared to detectors from the 7.5\,nm film. Here, we suggest two practical solutions to solve the trade-off between film thickness and performance for future mid-infrared SNSPDs study: Bias-assisted sputtering can be applied to improve the critical current of SNSPDs \cite{bias-assistant} and post-processing treatment (for example, helium ion irradiation \cite{post}) can enhance the internal efficiency of SNSPDs made from thicker films.


\section{Measurements of Fiber-coupled SNSPDs }

For most quantum optics experiments and applications, a fiber-coupled detector/system is preferred because of mature fiber optics technology and instruments.

\begin{figure}[!ht] 
    \centering
    \includegraphics[width=17cm]{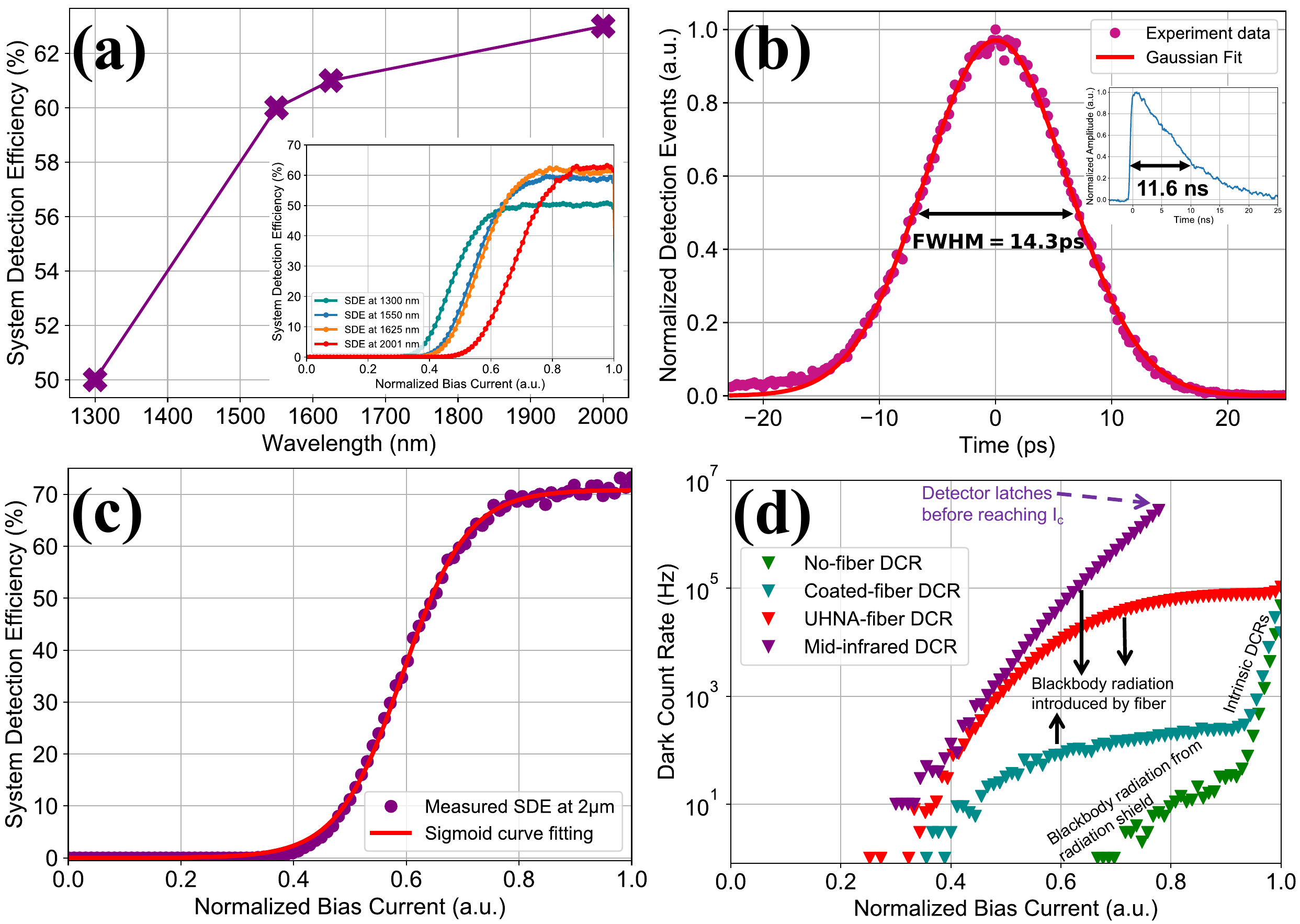}
    \caption{Fiber-coupled SNSPDs measurements of \textbf{(a)} SDE of detector \#1 at 1310, 1625, 1550 and 2000\,nm, \textbf{ (b)} timing jitter of detector \#1, \textbf{(c)} SDE of detector \#2 at 2001\,nm, and \textbf{(d)} DCRs of detector \#2 under different conditions: no fiber (green), low-pass coated fiber (turquoise), UHNA  fiber (red), and mid-IR fiber (purple) plugged in.}
    \label{fig4}
\end{figure}

The previous section provides evidence that both 7.5 and 9.5\,nm NbTiN superconducting films are suitable for making mid-infrared SNSPDs in terms of good yield and internal efficiency while a reduced thickness (5-6.5\,nm) leads to fewer working devices. Thus, we fabricated fiber-coupled SNSPDs from 7.5 \,nm thick NbTiN film. Similar to \cite{99delft}, the NbTiN films were initially deposited on SiO$_2$ grown by thermal oxidation process and nanowire meanders were eventually located on top of a Au/SiO$_2$ membrane acting as optical cavity. After packaging and wire bonding, detectors were mounted in a closed-cycle cryocooler with a base temperature of 2.4-2.8\,K and coupled to single mode fibers. Afterwards, lasers with different wavelengths were used for system detection efficiency (SDE) measurements as described in the previous section.  \par

Figure \ref{fig4} (a) shows the performance of detector \#1 (60\,nm line width) made from a 7.5\,nm NbTiN film. The SDEs for 1300, 1550, 1625 and 2000\,nm are 50\%, 60\%, 61\%, and 63\% respectively. The inset in figure \ref{fig4} (a) shows detector \#1's photon counting rate (PCR) curves at several wavelengths. Besides high SDE, high timing resolution is also highly desirable for many applications, for example, LiDAR \cite{2.3lidar}, fluorescence microscopy and spectroscopy \cite{mir-fluorescence,mir-spec}. The Instrument Response Function (IRF) of detector \#1 was characterized with a ps-pulsed laser (4.2\,ps pulse-width at 1064\,nm wavelength) and a fast oscilloscope (4\,GHz bandwidth, 40\,GHz sampling rate) as described in \cite{99delft}. As shown in figure \ref{fig4} (b), using a low-noise cryogenic amplifier operated at 40\,K, the IRF of this device shows a Gaussian shaped histogram. After fitting, we obtain 14.3$\pm$0.1\,ps timing jitter (full width at half maximum, FWHM). Compared to previous reported values for mid-infrared SNSPDs \cite{1.5-6um}, we improved time resolution by nearly two orders of magnitude. The inset picture in figure \ref{fig4} (b) shows the pulse trace of detector \#1. It shows a dead-time of 11.6\,ns, indicating good performance at high count rates \cite{esmaeil2017single}. Similarly, in figure \ref{fig4} (c), we show the 2000\,nm SDE measurement of detector \#2, which is made from another 7.5\,nm NbTiN film but has a slightly higher meander filling factor (approximately 10\% higher). The benefit of  a higher filling factor is an increased optical absorption, which means if the internal efficiency is saturated, a higher SDE can be achieved compared to a similar device with a lower filling factor. As can be seen, detector \#2 shows well saturated internal efficiency (when I$_b$ $\geqslant$ 0.8I$_c$). After sigmoid fitting (to improve the efficiency estimation accuracy), we obtained a peak SDE over 70\% at 2000\,nm.\par

Besides achieving high system detection efficiency, high dark count rates of mid-infrared SNSPDs are a major challenge. Previous work \cite{2.3lidar} showed that the DCRs of mid-infrared SNSPDs is typically in the order of $10^4$ Hz without using additional filters. In figure \ref{fig4} (d), we systematically studied the DCRs of detector \#2 in four different schemes: DCR without any fiber connected to the detector (green curve), DCR with end-face coated SM2000 fiber (fiber operating wavelength 1.7-2.3 \textmu m, cyan curve), DCR with ultra-high NA fiber without coating (fiber operating wavelength 1.5-2\,\textmu m, red curve) and DCR with mid-infrared ZrF$_4$ fiber without coating (fiber operating wavelength 2.3-4.1 \textmu m, purple curve). As a result, at I$_b$=0.8I$_c$, the DCRs of detector \#2 for the above mentioned four schemes are around the order of 10$^1$ Hz, 10$^2$ Hz, 10$^5$ Hz and 10$^6$ Hz, respectively. It is clear that the DCR of detector \#2 when coupled to an end-face coated fiber is 3 orders of magnitude lower than coupled to ultra-high NA fiber without coating. By using this fiber end-facet coating (low-pass filter), the DCR of detector \#2 is below 240\,Hz when it reaches unity internal efficiency at 2000 nm. In contrast, when detector \#2 is connected to mid-infrared ZrF$_4$ fiber for SDE measurements at 3-4 \textmu m, the detector showed over 2.78\,MHz DCR at 0.8 I$_c$ bias and starts latching \cite{latching}. This prevents further SDE measurements of detector \#2 at 3-4 \textmu m. To solve this issue, low-pass filters should be employed before the detectors, either by using fiber end-face coating, or adding cold filtering stages inside the cryostat \cite{cold-filters}.

\section{Discussion and Conclusion}

In the past, amorphous materials were mainly used for mid-infrared single photon detection motivated by the intuition that their superconducting energy gap (0.59-0.61\,meV for WSi \cite{zhang2016characteristics}) is lower than polycrystalline material (2.46 \,meV for NbN \cite{NbNgap}). This work pinpoints that NbTiN (polycrystalline) based SNSPDs can also achieve high mid-infrared single photon detection efficiency while maintaining unprecedented time resolution. Furthermore, given that the energy of a single photon even at 10\,\textmu m wavelength (123.9\,meV) is still significantly larger than both materials' superconducting energy gap, other physical properties of the superconducting materials need to be investigated to enhance SNSPDs' mid-infrared detection response. Besides improving internal detection efficiency, reducing the dark count rates is another outstanding challenge for mid-infrared SNSPD systems. As shown in this work, only room temperature black body radiation delivered to the detector by ZrF$_4$ fiber has led to >10$^6$\,Hz DCR. To overcome this issue, either extra cryogenic filters need to be added before the detectors or the entire experiment has to be performed at cryogenic temperatures.        \par
In conclusion, we demonstrated SNSPDs made from magnetron co-sputtered NbTiN superconducting films (5-9.5\,nm) with unity internal efficiency at 3\,\textmu m and 80\% internal efficiency at 4013\,nm when operated in closed-cycle Gifford-McMahon coolers (2.4-2.8\,K). Our fiber coupled device achieves over 70\% system detection efficiency at 2 \textmu m and > 50\% system detection efficiency from 1300 to 2000\,nm with sub-15\,ps time resolution. By employing an end-facet coated fiber, the dark count rate of mid-infrared SNSPDs was reduced by 3 orders of magnitude compared to uncoated single mode fibers. The DCR when coupled to mid-infrared ZrF$_4$ fiber is also studied, which offers valuable information for building mid-infrared fiber-coupled SNSPDs systems in the future. To the best of our knowledge, the detectors presented in this work have the best system detection efficiency and temporal resolution among the mid-infrared SNSPDs reported so far, and NbTiN is a solid choice for making mid-infrared SNSPDs without mK dilution refrigerators.

\section*{Acknowledgements}

J.C. acknowledges China Scholarships Council (CSC, No.201603170247) . I.E.Z., V.Z., and Single Quantum B.V. acknowledge the supports from the ATTRACT project funded by the EC under Grant Agreement 777222. R.B.M.G. acknowledges support by the European Commission via the Marie-Sklodowska Curie action Phonsi (H2020-MSCA-ITN-642656). S.N.D., S.S., V.Z. and Single Quantum B.V. acknowledge EU FET-Open project funding (No. 899580). V.Z. acknowledges funding from the Knut and Alice Wallenberg Foundation Grant “Quantum Sensors”, and support from the Swedish Research Council (VR) through the VR Grant for International Recruitment of Leading Researchers (Ref 2013-7152) and Research Environment Grant (Ref 2016-06122).

\bibliography{sample.bib}
\end{document}